\documentclass[12pt,oneside,reqno]{amsart}
\usepackage[a4paper]{geometry}
\usepackage[T1]{fontenc}
\usepackage[latin9]{inputenc}
\usepackage[all, warning]{onlyamsmath}
\usepackage{amstext}
\usepackage{amssymb}
\usepackage{natbib}

\usepackage{graphicx}
\RequirePackage[l2tabu, orthodox]{nag}
\usepackage[subrefformat=parens]{subcaption}

\DeclareMathOperator*{\argmin}{arg\,min}
\newcommand{\K}[2]{K\left(\frac{#2}{#1}\right)}
\newcommand{\E}[2][]{\mathbb{E}_{#1}\mathopen{}\left[#2\mathopen{}\right]}
\newcommand{\Ec}[3][]{\mathbb{E}_{#1}\mathopen{}\left[#2 \middle| #3\mathopen{}\right]}
\newcommand{\converge}[1]{\stackrel{#1}{\longrightarrow}}
\newcommand{\set}[1]{\left\{#1\right\}}

\begin{document}
\title[Semiparametric estimation of heterogeneous treatment effects]{Semiparametric estimation of heterogeneous treatment effects under the nonignorable assignment condition}
\author{Keisuke Takahata$^{1),2)}$}
\author{Takahiro Hoshino$^{1),2)}$}
\thanks{1) Department of Economics, Keio University, Tokyo, Japan; 2) RIKEN Center for Advanced Intelligence Project, Tokyo, Japan}
\begin{abstract}
We propose a semiparametric two-stage least square estimator for the heterogeneous treatment effects (HTE). HTE is the solution to certain integral equation which belongs to the class of Fredholm integral equations of the first kind, which is known to be ill-posed problem. Naive semi/nonparametric methods do not provide stable solution to such problems. Then we propose to approximate the function of interest by orthogonal series under the constraint which makes the inverse mapping of integral to be continuous and eliminates the ill-posedness. We illustrate the performance of the proposed estimator through simulation experiments. 

\vspace*{2pt}
\end{abstract}
\maketitle

\section{Introduction}
In causal inference, treatment effects such as average treatment effect (ATE) or average treatment effect on the treated (ATT) have been of primary interest in the literature ~\citep{rubin_estimating_1974}. These parameters are, as the names stand for, an \textit{averaged} treatment effect over a population of interest. However, a treatment effect may differ among units depending on the covariates and outcomes. Such heterogeneity of treatment effects has been intensively studied in recent years. For example, \cite{wager_estimation_2018} proposed a method for finding subgroups in which the treatment effect is similar using the random forest. Understanding heterogeneity of treatment effects aids not only to more detailed analysis of a population of interest but also to a more efficient policy-making where an intervention is costly.

While most studies have concerned heterogeneity over the covariates which are fully observed, \cite{takahata_identification_2018} (henceforth TH) studied the heterogeneity over the untreated potential outcome, which is defined as
\begin{align}
\mathrm{HTE}(y_{0}) = \Ec{y_{1}-y_{0}}{y_{0}},
\label{eq:hte}
\end{align}
where $y_{1}$ and $y_{0}$ are the outcome when receiving the treatment and control condition respectively. Following TH, we call (\ref{eq:hte}) as the heterogeneous treatment effect (HTE), which is also of interest in this paper. To estimate the HTE, we have to deal with non-ignorable missingness because we need to estimate $p(y_{1}|y_{0})$ but $y_{1}$ and $y_{0}$ are never observed simultaneously. It is known that identification is not trivial in non-ignorable missing models (e.g., \cite{miao_identifiability_2016}). TH provided the sufficient condition for the identification of the HTE using the information of the marginal distribution of $y_{0}$, $p(y_{0})$. Although the identification condition is rather general, estimation of the HTE is difficult in that it is necessary to solve some integral equation; the integral equation that we need to consider is a Fredholm integral equation of the first kind, which is known to be a ill-posed problem. In general, appropriate regularization methods are needed to obtain a stable solution to such equation. In TH, this problem was avoided using a parametric Bayesian modeling, but for wide applicability, a more flexible approaches is desired. 

In this paper, we propose a semiparametric two-stage least square (2SLS) estimator for the HTE. Our approach relies on the quadratic programming method proposed by \cite{newey_instrumental_2003}, in which they concerned the estimation of a nonparametric instrumental variable model. The function of interest is approximated by series of a finite number and then the integral equation reduces to a constrained least square problem with the regressors replaced by its expectation. To overcome instability of the solution due to the ill-posedness, certain bounds are placed on the coefficients of the series to make the inverse mapping of integral to be continuous. The numerical experiments show that the proposed method correctly estimate the HTE. 

\section{Model setup}
We follow the same setup as TH. The HTE (eq.~(\ref{eq:hte})) is rewritten as
\begin{align*}
\Ec{y_{1}-y_{0}}{y_{0}} = \Ec{y_{1}}{y_{0}} - y_{0} = \E[x|y_{0}]{\Ec{y_{1}}{y_{0},x}} - y_{0},
\end{align*}
where $x\in \mathbb{R}^{d}$ is $d$-dimensional covariate. From this formula we observe that, for the identification of the HTE, it is sufficient to identify $p(x|y_{0})$ and $p(y_{1}|y_{0},x)$. Let $z\in\set{0,1}$ be the binary indicator, which is equal to 1 when assigned to the treatment condition. If $z=1$, then only $y_{1}$ is observed and $y_{0}$ is missing, and vice versa. TH showed that the following two assumptions play a primary role for the identification:
\begin{description}
\item[(A1)] $p(z|y_{1},y_{0},x)=p(z|y_{0},x)$ (\textit{weak ignorability});
\item[(A2)] $p(y_{0})$ is known.
\end{description}
Assumption (A1) is called weak ignorability. Intuitively, this assumption implies that we can identify the HTE by observing the difference of two groups in which the assignment probability depends on $y_{0}$. Therefore, in a situation where strong ignorability, $p(z|y_{1},y_{0},x)=p(z|x)$, is satisfied, our approach is not applicable. Assumption (A2) is needed for identify $p(z|y_{0},x)$ \citep{hirano_combining_2001}. In addition to these conditions, several constraints on the functional form and parameter space of $p(z|y_{0},x)$ is needed; for more detailed discussion on the identification condition, see TH.  In what follows, we suppose that all the conditions mentioned in Theorem 2 in TH are satisfied. Note that, in this setup, the identification of ATE is trivial because
\begin{align*}
\E{y_{1}-y_{0}} = \Ec[y_{0},x]{y_{1}}{y_{0},x} - \E{y_{0}}.
\end{align*}

Consider the integral equation
\begin{align}
\notag \E{y_{1}|x,z=1} 
&= \int \E{y_{1}|y_{0},x,z=1}p(y_{0}|x,z=1)dy_{0} \\
&= \int \E{y_{1}|y_{0},x}p(y_{0}|x,z=1)dy_{0},
\label{eq:integral_eq}
\end{align}
where the second equality holds by weak ignorability. Under the identification condition, it can be proved that the solution to eq.~(\ref{eq:integral_eq}) for $\E{y_{1}|y_{0},x}$ is unique, that is, $\E{y_{1}|y_{0},x}$ is identified. Then our goal is to obtain an actual solution to eq.~(\ref{eq:integral_eq}). However, if we employ a nonparametric method for estimating $\Ec{y_{1}}{y_{0},x}$, a solution suffers from the instability due to the discontinuity of the inverse mapping of integral. Then we need to take an appropriate regularization method to overcome the ill-posedness of eq.~(\ref{eq:integral_eq}). We address this problem in the next section. 

\section{Estimation}
In this section we propose a two-stage least square estimator (2SLSE) for $\Ec{y_{1}}{y_{0},x}$ based on \cite{newey_instrumental_2003}'s method. The strategy is that (i) we approximate $\Ec{y_{1}}{y_{0},x}$ by a finite number of orthogonal basis functions, (ii) take expectation of them with respect to $p(y_{0}|x,z=1)$, and (iii) do least square estimation under the constraint to make the inverse mapping of the integral to be continuous and eliminate the ill-posedness of eq.~(\ref{eq:integral_eq}). For simplicity we suppose $x\in\mathbb{R}$ in the rest of the paper. 

\subsection{Estimation of $\E{y_{1}|y_{0},x}$}
We consider approximating $\phi(y_{0},x)\equiv \Ec{y_{1}}{y_{0},x}$ with a finite number of orthogonal basis functions, $\{q_{j}\}_{j=0}^{J}$, as
\begin{align}
\phi(y_{0},x) \simeq \sum_{j_{1}=0}^{J}\sum_{j_{2}=0}^{J} \gamma_{j_{1} j_{2}}q_{j_{1}}(y_{0})q_{j_{2}}(x).
\label{eq:approx}
\end{align}
We specify $p(z=1|y_{0},x)$ by the logistic regression such that it satisfies the identification condition from TH:
\begin{align}
g(k_{0}+k_{y_{0}}(y_{0}) + k_{x}(x)) = \frac{1}{1+\exp(-(k_{0}+k_{y_{0}}(y_{0}) + k_{x}(x)))},
\label{eq:missing_mechanism}
\end{align}
where the additivity holds between $y_{0}$ and $x$. Expand $p(y_{0}|x,z=1)$ as
\begin{align}
\notag
p(y_{0}|x,z=1) &= \frac{p(z=1|y_{0},x)}{p(z=0|y_{0},x)}\frac{p(z=0)}{p(x,z=1)}p(y_{0}x|z=0) \\
\notag
&=\left\{\frac{ \exp(k_{0} + k_{x}(x))p(z=0)}{p(x,z=1)}\right\}\exp(k_{y_{0}}(y_{0}))p(y_{0},x|z=0)\\
&= c(x) \exp(k_{y_{0}}(y_{0}))p(y_{0},x|z=0), 
\label{eq:p_y0_xz1_re}
\end{align}
where $c(x) = \exp(k_{0} + k_{x}(x))p(z=0)/p(x,z=1)$.
Plugging eq.~(\ref{eq:approx}) and eq.~(\ref{eq:p_y0_xz1_re})
into eq.~(\ref{eq:integral_eq}) yields
\begin{align}
\notag &E[y_{1}|x,z=1] \\
\notag \simeq &\int \left[\sum_{j_{1}=0}^{J}\sum_{j_{2}=0}^{J} \gamma_{j_{1} j_{2}}q_{j_{1}}(y_{0})q_{j_{2}}(x)\right]\hat{c}(x) \exp(\hat{k}_{y_{0}}(y_{0}))p(y_{0},x|z=0) dy_{0} \\
\notag = &\sum_{j_{1}=0}^{J}\sum_{j_{2}=0}^{J} \gamma_{j_{1} j_{2}}c(x)q_{j_{2}}(x)\int q_{j_{1}}(y_{0})\exp(k_{y_{0}}(y_{0}))p(y_{0},x|z=0)dy_{0} \\
= &\sum_{j_{1}=0}^{J}\sum_{j_{2}=0}^{J} \gamma_{j_{1} j_{2}}c(x)s_{j_{1}}(x)q_{j_{2}}(x),\label{eq:integral_approx}
\end{align}
where $s_{j_{1}}(x) = \int q_{j_{1}}(y_{0})\exp(k_{y_{0}}(y_{0}))p(y_{0},x|z=0)dy_{0}$.  The estimation of the missing mechanism will be discussed later.

Here, we consider estimating $p(y_{0},x|z=0)$ by the kernel density estimator, 
\begin{align*}
\hat{p}(y_{0},x|z=0) = \frac{1}{N_{0}}\sum_{i:z_{i}=0}\frac{1}{h_{y_{0}}h_{x}}\K{h_{y_{0}}}{y_{0}-y_{i0}}\K{h_{x}}{x-x_{i}}.
\end{align*}
where $h_{y_{0}}$ and $h_{x}$ are the bandwidths and $N_{0}$ is the sample size of the control group.
In this case, $s_{j_{1}}(x)$ can be estimated by
\begin{align}
\notag\hat{s}_{j_{1}}(x) 
&= \frac{1}{N_{0}}\sum_{i:z_{i}=0}\frac{1}{\hat{h}_{y_{0}}\hat{h}_{x}}\K{\hat{h}_{x}}{x-x_{i}}\left[\int q_{j_{1}}(y_{0})\exp(\hat{k}_{y_{0}}(y_{0}))\K{\hat{h}_{y_{0}}}{y_{0}-y_{i0}}dy_{0}\right]\\
&= \frac{1}{N_{0}}\sum_{i:z_{i}=0}\frac{1}{\hat{h}_{y_{0}}\hat{h}_{x}}\K{\hat{h}_{x}}{x-x_{i}}\hat{t}_{j_{1}}(y_{i0})
\label{eq:s_hat}
\end{align}
where $\hat{t}_{j_{1}}(y_{i0}) = \int q_{j_{1}}(y_{0})\exp(\hat{k}_{y_{0}}(y_{0}))\K{\hat{h}_{y_{0}}}{y_{0}-y_{i0}}dy_{0}$. Similarly, we obtain an estimator for $c(x)$, $\hat{c}(x)$, by the kernel density estimator,
\begin{align}
\hat{p}(x|z=1) = \frac{1}{N_{1}}\sum_{i:z_{i}=1}\frac{1}{\hat{w}_{x}}\K{\hat{w}_{x}}{x-x_{i}}.
\label{eq:kde_x_z1}
\end{align}

By inserting eq.~(\ref{eq:s_hat}) and eq.~(\ref{eq:kde_x_z1}) into eq.~(\ref{eq:integral_approx}), we obtain
\begin{align}
\notag &E[y_{1}|x,z=1] 
= \hat{c}(x)\sum_{j_{1}=0}^{J}\sum_{j_{2}=0}^{J} \gamma_{j_{1} j_{2}}\hat{s}_{j_{1}}(x) q_{j_{2}}(x),
\end{align}
where $N_{1}$ is the sample size of the treatment group. Therefore, the least square estimator for $\gamma_{j_{1} j_{2}}$ is obtained by the following quadratic problem:
\begin{align*}
\hat{\gamma}_{j_{1} j_{2}} 
&= \argmin_{\gamma_{j_{1} j_{2}}} \frac{1}{N_{1}}\sum_{i: z_{i}=1}\left( y_{i1}- \hat{c}(x_{i})\sum_{j_{1}=0}^{J}\sum_{j_{2}=0}^{J} \gamma_{j_{1} j_{2}}\hat{s}_{j_{1}}(x_{i}) q_{j_{2}}(x_{i}) \right)^{2},
\end{align*}
\begin{align*}
\text{s.t.}\quad\quad \gamma'\Lambda_{J} \gamma \,\leq\, B_{\gamma},
\end{align*}
where $\gamma = (\gamma_{11},\gamma_{12},\dots,\gamma_{j_{1}j_{2}},\dots,\gamma_{JJ})'$,
 $B_{\gamma}$ is a positive constant. $\Lambda_{J}$ is the Sobolev norm of the basis functions, $\set{q_{j_{1}}(\cdot)q_{j_{2}}(\cdot)}_{(j_{1},j_{2})}\, (j_{1}=0,\dots,J, \, j_{2}=0,\dots,J)$, which imposes compactness on both the true and estimated functions of $\phi(y_{0},x)$. This compactness eliminates the ill-posedness of the inverse problem (eq.~(\ref{eq:integral_eq})) because the inverse operator of the integral becomes continuous mapping \citep{newey_instrumental_2003}.

Finally, by integrating out $x$ in $\hat{\phi}$, we obtain $\hat{E}[y_{1}|y_{0}]$:
\begin{align*}
\hat{E}[y_{1}|y_{0}] &= \int \hat{E}[y_{1}|y_{0},x]\hat{p}(x|y_{0}) dx \\
&= \int \hat{\phi}(y_{0},x)\hat{p}(x|y_{0}) dx.
\end{align*}
Note that we can calculate $\hat{p}(x|y_{0})$ by plugging corresponding estimators into the following formula:
\begin{align*}
p(x|y_{0}) = \frac{p(y_{0},x)}{p(y_{0})} = \frac{p(y_{0},x|z=0)p(z=0)}{p(z=0|y_{0},x)p(y_{0})}.
\end{align*}

\subsection{Estimation of the missing mechanism}
We estimate the missing mechanism (eq.~(\ref{eq:missing_mechanism})) referring to \cite{nevo_using_2003}, who proposed a generalized method of moments (GMM) estimator for a nonignorable missing model. Suppose that the auxiliary moment condition, $\E{m(x,y_{0})}=0$, is available. From assumption (A2), we can calculate any moments of $y_{0}$ up to infinite dimension, but for simplicity, the dimension of the moment condition is set to be equal to the sum of the dimension of the parameters of $k_{y_{0}}(\cdot)$ and $k_{x}(\cdot)$. For example, if $k_{x}(x)=\beta_{0}x$ and $k_{y_{0}}(y_{0})=\beta_{1}y_{0}+\beta_{2}y_{0}^{2}$, then we may set the moment function as
\begin{align*}
m(x,y_{0})=(x-\E{x},\, y_{0}-\E{y_{0}},y_{0}^{2}-\E{y_{0}^{2}})'.
\end{align*}
Note that
\begin{align*}
\Ec{\frac{m(x,y_{0})}{p(z=0|y_{0},x)}}{z=0} 
&= \int \frac{m(x,y_{0})}{p(z=0|y_{0},x)}p(y_{0},x|z=0)dy_{0} dx\\
&= p(z=0)\int m(x,y_{0})p(y_{0},x)dy_{0} dx\\
&= p(z=0)\E{m(x,y_{0})}=0.
\end{align*}
Therefore, the solution of the following system of equations,
\begin{align*}
&\frac{1}{N_{0}}\sum_{i:z_{i}=0} m(x_{i},y_{i0})(1+\exp(k_{0}+k_{x}(x_{i})+k_{y_{0}}(y_{i0}))) = 0\\
&\sum_{i:z_{i}=0} (1+\exp(k_{0}+k_{x}(x_{i})+k_{y_{0}}(y_{i0}))) = N,
\end{align*}
is an unbiased estimator for $(k_{0},\beta_{0},\beta_{1},\beta_{2})$. $N$ is the total sample size. The second equation implies the normalization of the weights, which is due to $1/N\cdot \sum_{i:z_{i}=0}1/p(z_{i}=0|y_{i0},x_{i})\converge{p}1$. For detailed identification conditions, see \cite{nevo_using_2003}.

\section{Simulation}
\begin{figure}[htbp]
	\centering
	\includegraphics[width=0.5\linewidth]{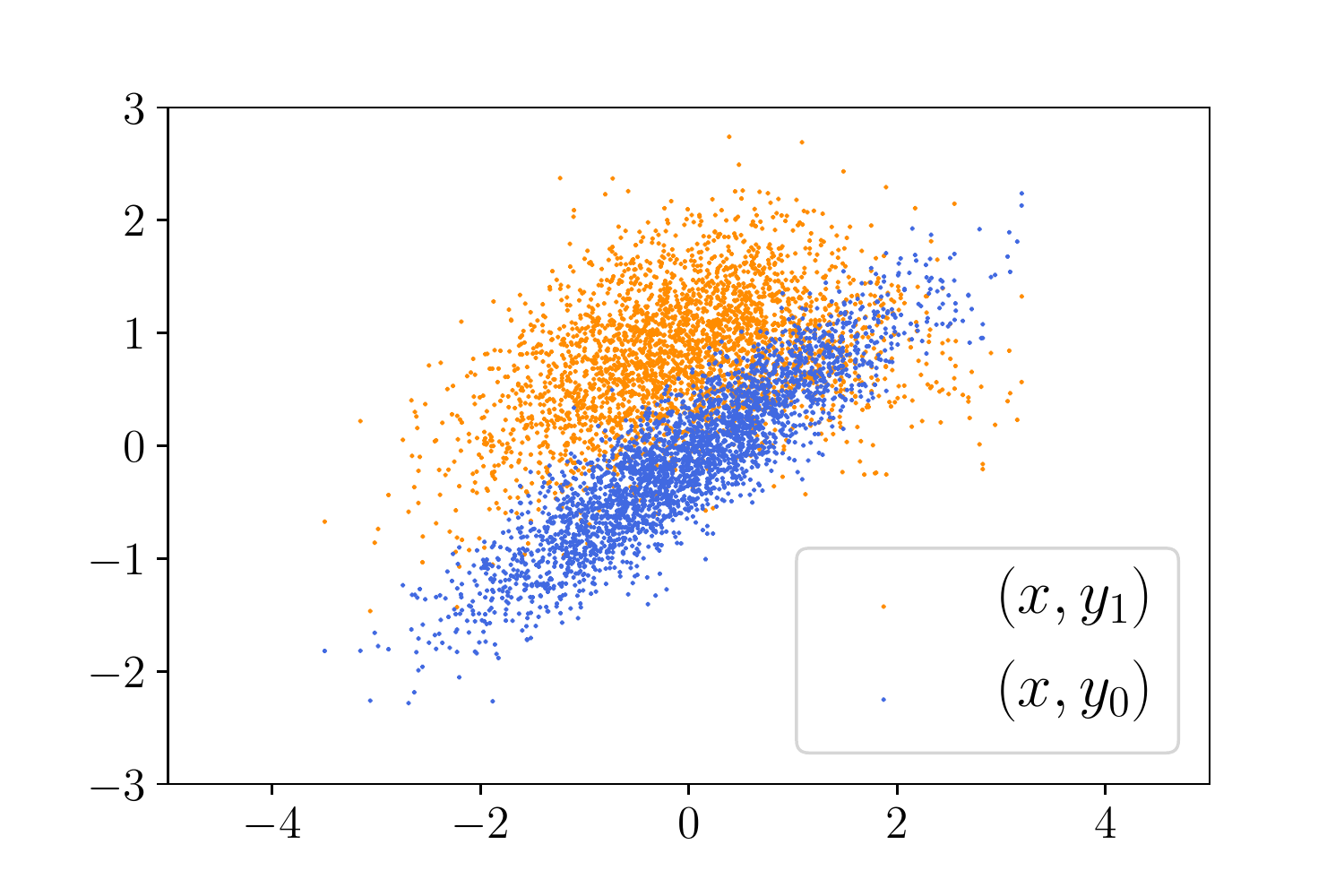}
	\caption{The scatter plot of the complete data: $(x,y_{0})$ and $(x,y_{1})$}
	\label{fig:plots_xy} 
\end{figure}

\begin{figure}[htbp]
	\centering
	\subcaptionbox{$(x,y_{0})|z=0$ (observed)\label{fig:plot_x_y0_obs}}
	{\includegraphics[width=0.40\linewidth]{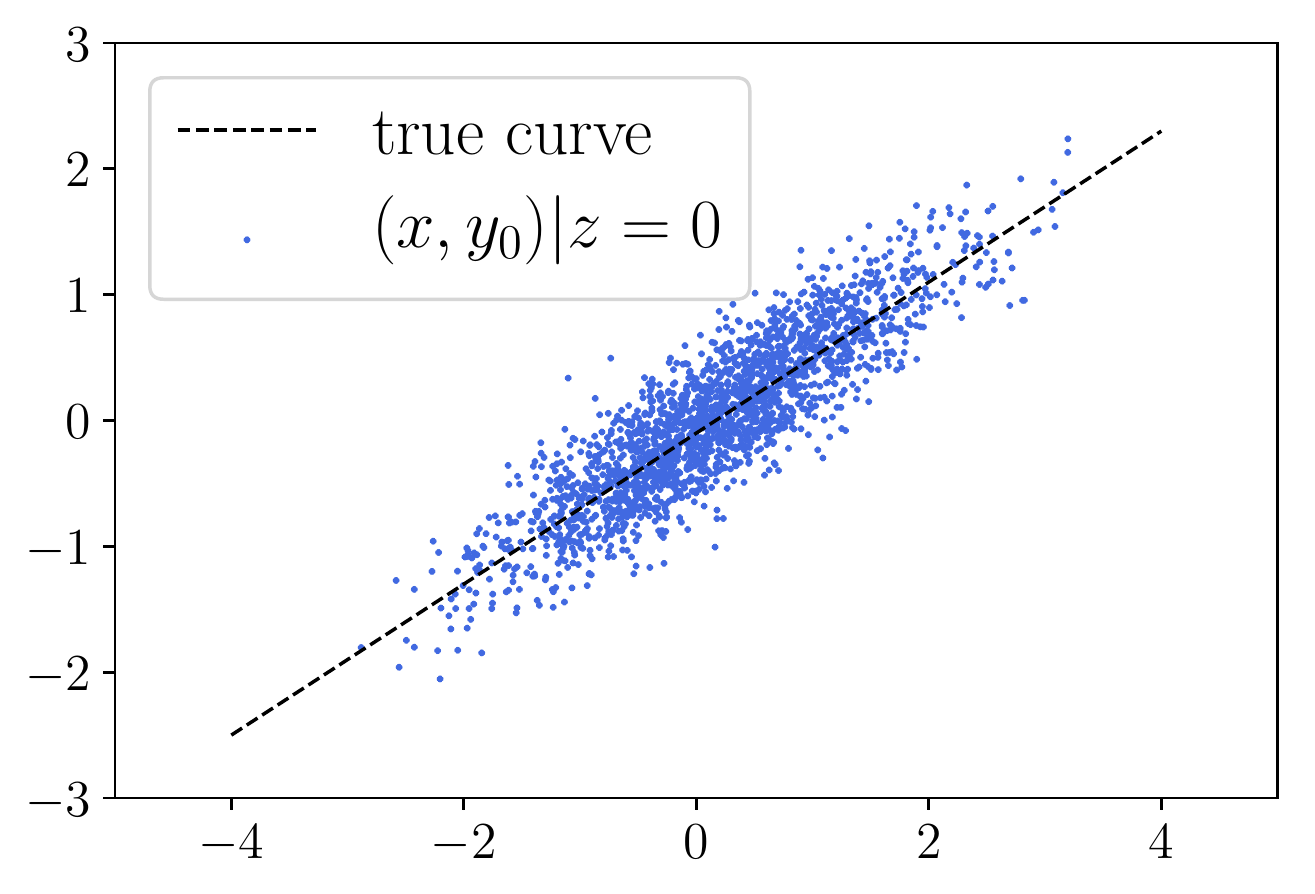}}
	\subcaptionbox{$(x,y_{1})|z=1$ (obserbed)\label{fig:plot_x_y1_obs}}
	{\includegraphics[width=0.40\linewidth]{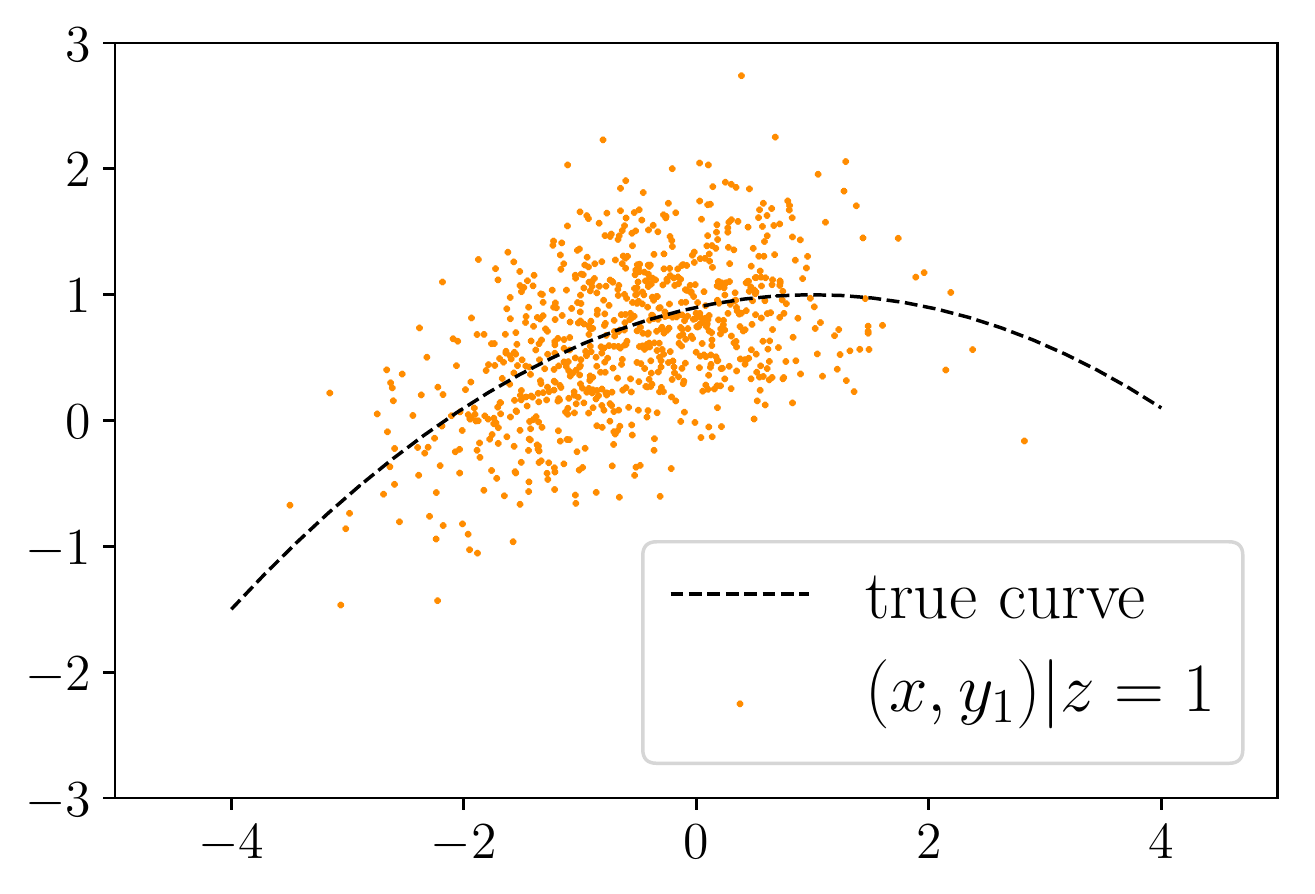}}	
	\vspace{2mm} \\
	\subcaptionbox{$(x,y_{0})|z=1$ (missing)\label{fig:plot_x_y0_mis}}
	{\includegraphics[width=0.40\linewidth]{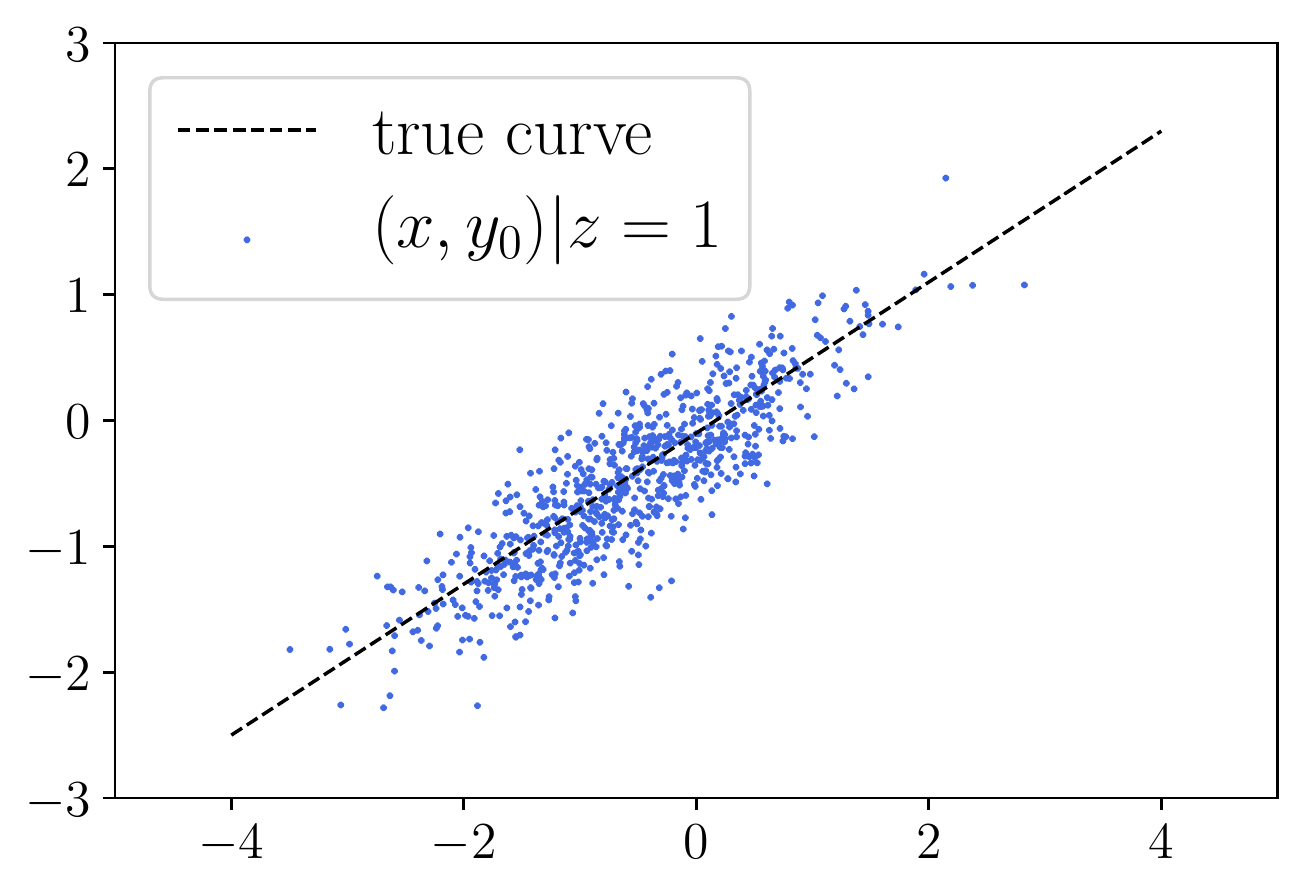}}
	\subcaptionbox{$(x,y_{1})|z=0$ (missing)\label{fig:plot_x_y1_mis}}
	{\includegraphics[width=0.40\linewidth]{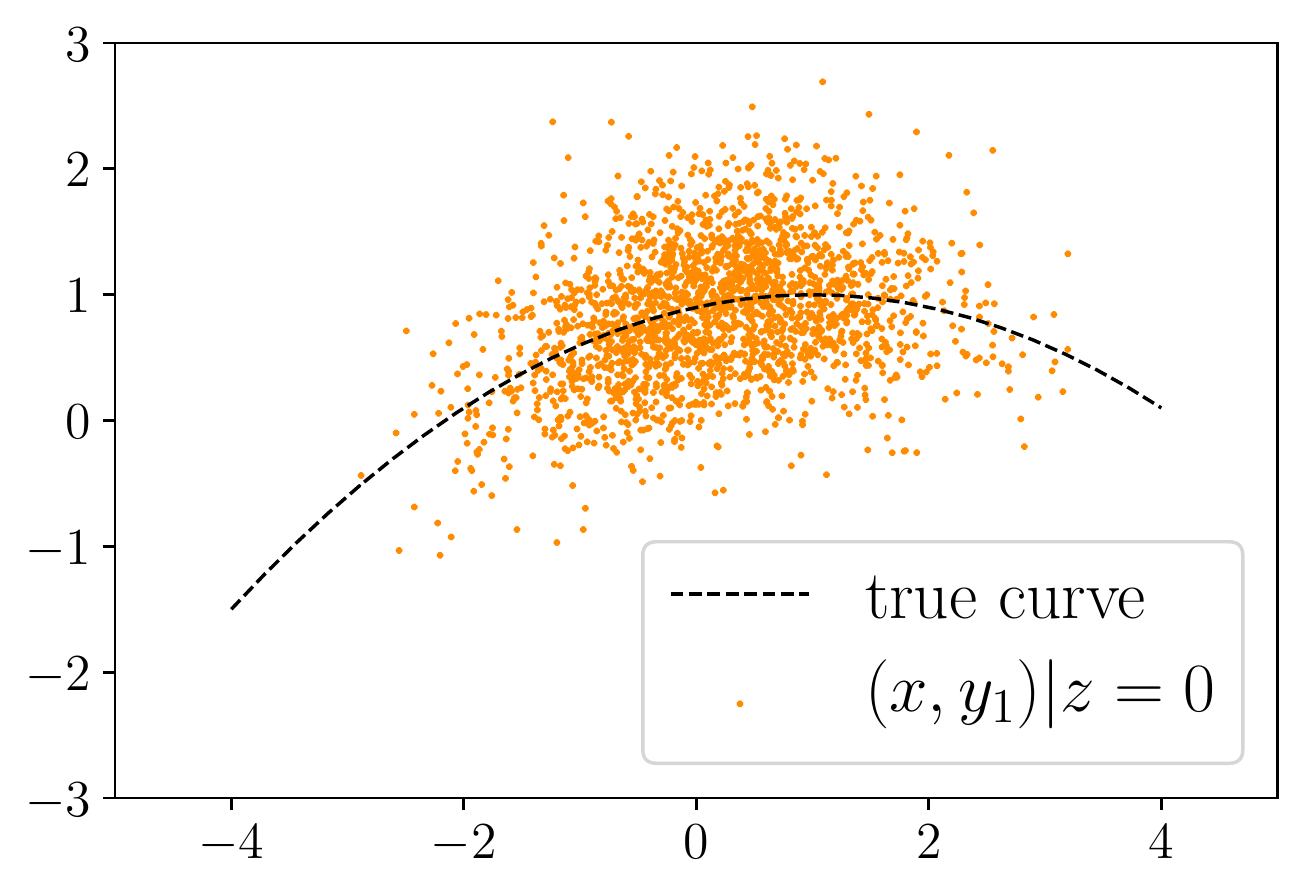}}
	\caption{Comparison between the treatment and the control group}
	\label{fig:plots_xy_comp} 
\end{figure}

We conduct simulations to examine the performance of the estimator shown in the previous section. Each data set is generated from
\begin{align*}
&x \sim \mathrm{N}(0, 1), \quad
\left.\begin{pmatrix}
y_{0} \\
y_{1}
\end{pmatrix} \right| x
\sim \mathrm{N}\left(
\begin{pmatrix}
\mu_{0}(x) \\
\mu_{1}(x)
\end{pmatrix}, \quad
\begin{pmatrix}
\sigma_{0}^{2} &\rho\sigma_{0}\sigma_{1} \\
\rho\sigma_{0}\sigma_{1} &\sigma_{1}^{2}
\end{pmatrix}
\right)
\end{align*}
where $\sigma_{0} = 1/5$, $\sigma_{1} = 1/2$, $\rho = 1/2$, $\mu_{0}(x) = -3x/5 - 1/10$, $\mu_{1}(x) = -(x-1)^{2}/10 + 1$ and the sample size is $N=3000$ (Figure \ref{fig:plots_xy}, \ref{fig:plots_xy_comp}). In this study, we use Legendre polynomials up to the third-order ($J=3$) as the basis functions in eq.~(\ref{eq:approx}):
\begin{align*}
q_{0}(v) = 1, \quad q_{1}(v) = v, \quad q_{2}(v) = \frac{1}{2}(3v^{2}-1), \quad q_{3}(v) = \frac{1}{2}(5v^{3}-3v).
\end{align*}
Then, we make an appropriate linear transformation so that each variable is included in $[-1,1]$ because Legendre polynomials are orthogonal on this interval (hereafter, the variables denote the transformed values. After estimation, the inverse transformation is made to calculate ATE). We specify the functions in the missing mechanism (eq.~(\ref{eq:missing_mechanism})) as
$k_{x}(x) = \beta_{0} x$ and $k_{y_{0}}(y_{0}) = \beta_{1} y_{0} + \beta_{2}y_{0}^{2}$ 
and $z_{i}$ is drawn from $\mathrm{B}(1,p_{i})$ where $p_{i} = g(k_{0}+k_{y_{0}}(y_{i0}) + k_{x}(x_{i}))$. We set $k_{0} = -3/2$, $\beta_{0} = -2$, $\beta_{1} = -2$, $\beta_{2} = 1$ and the mean of the probability of being assigned to the treatment group becomes about $30\%$. For the units where $z_{i} = 1$, only $y_{i1}$ and $x_{i}$ are used for estimation, and conversely, $y_{i0}$ and $x_{i}$ are used where $z_{i} = 0$. Bandwidths for the kernel density estimators are chosen by Scott's normal reference rule of thumb. For estimation, we first estimate the missing mechanism, and then $\Ec{y_{1}}{y_{0},x}$ given the former estimator.

\begin{figure}[htbp]
	\centering
	\subcaptionbox{$B_{\gamma}=10$\label{fig:res_b10}}
	{\includegraphics[width=0.40\linewidth]{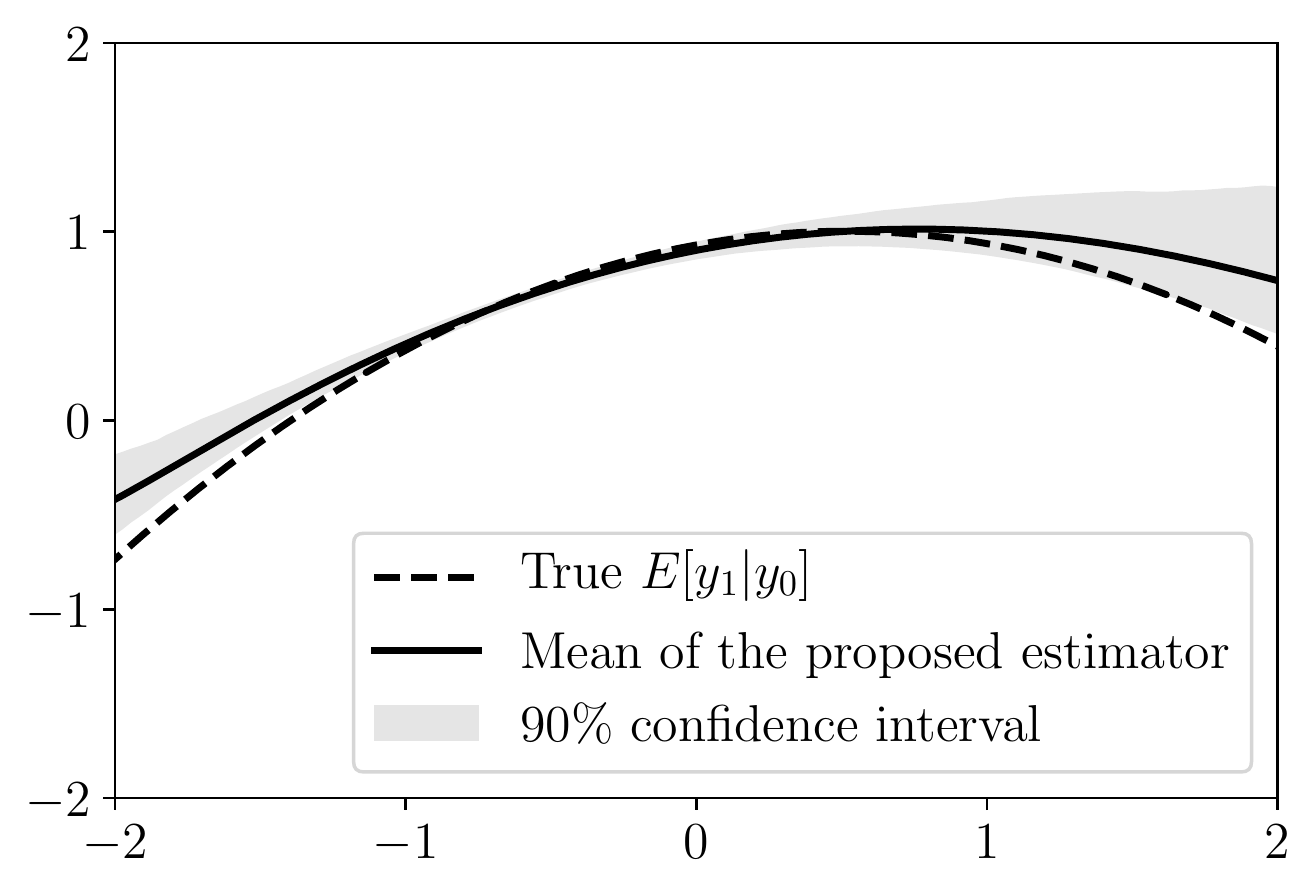}}
	\subcaptionbox{$B_{\gamma}=15$\label{fig:res_b15}}
	{\includegraphics[width=0.40\linewidth]{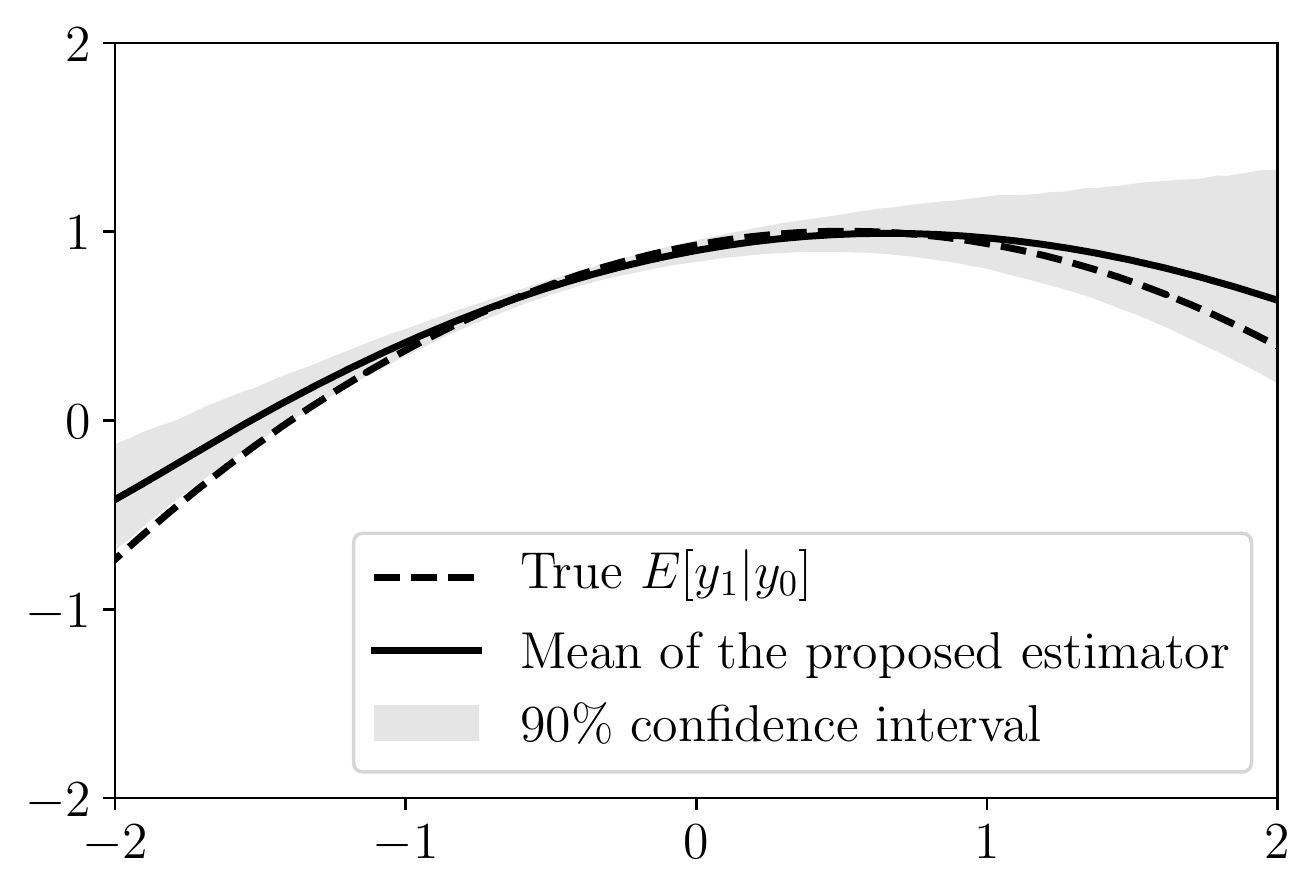}}
	\vspace{2mm} \\
	\subcaptionbox{$B_{\gamma}=25$\label{fig:res_b25}}
	{\includegraphics[width=0.40\linewidth]{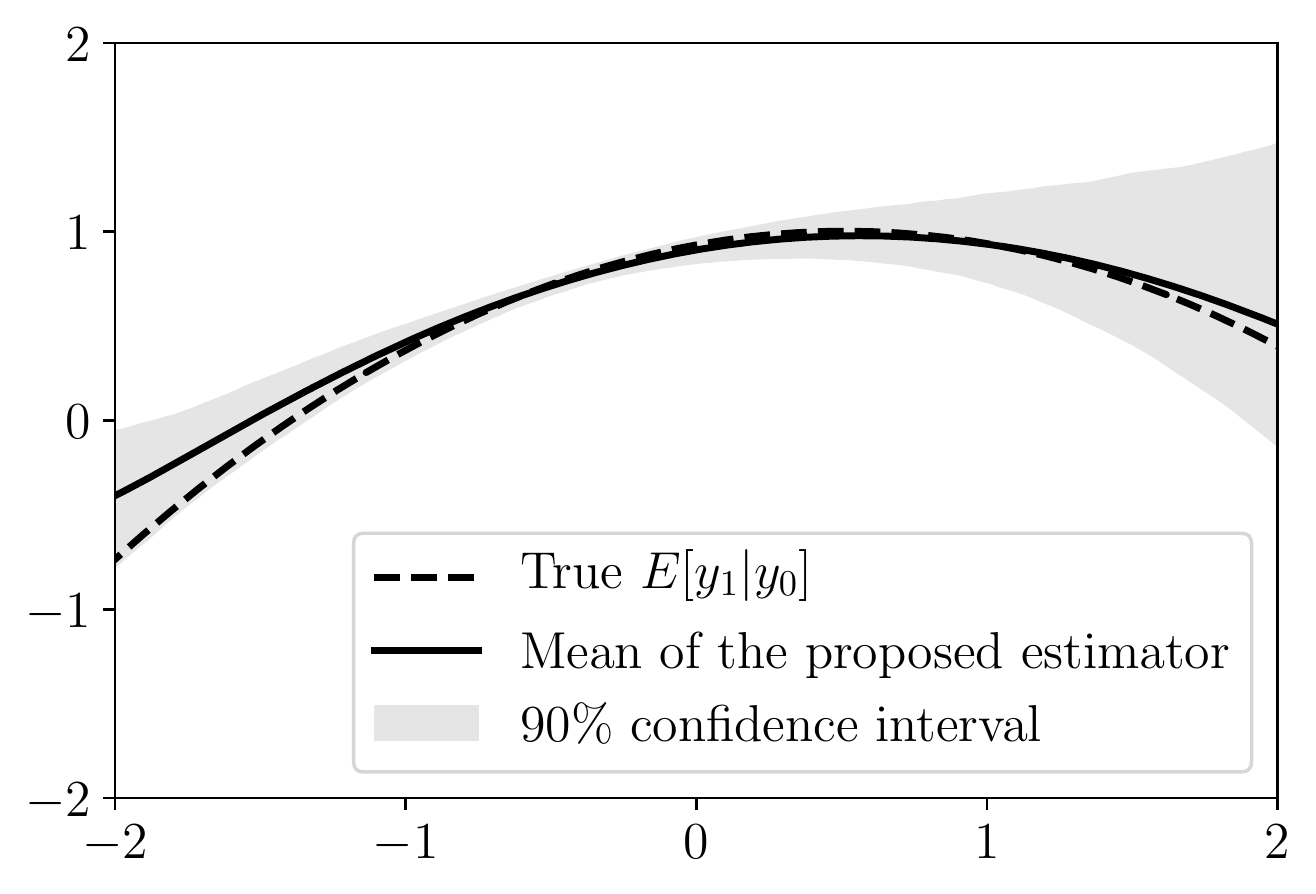}}
	\subcaptionbox{$B_{\gamma}=50$\label{fig:res_b50}}
	{\includegraphics[width=0.40\linewidth]{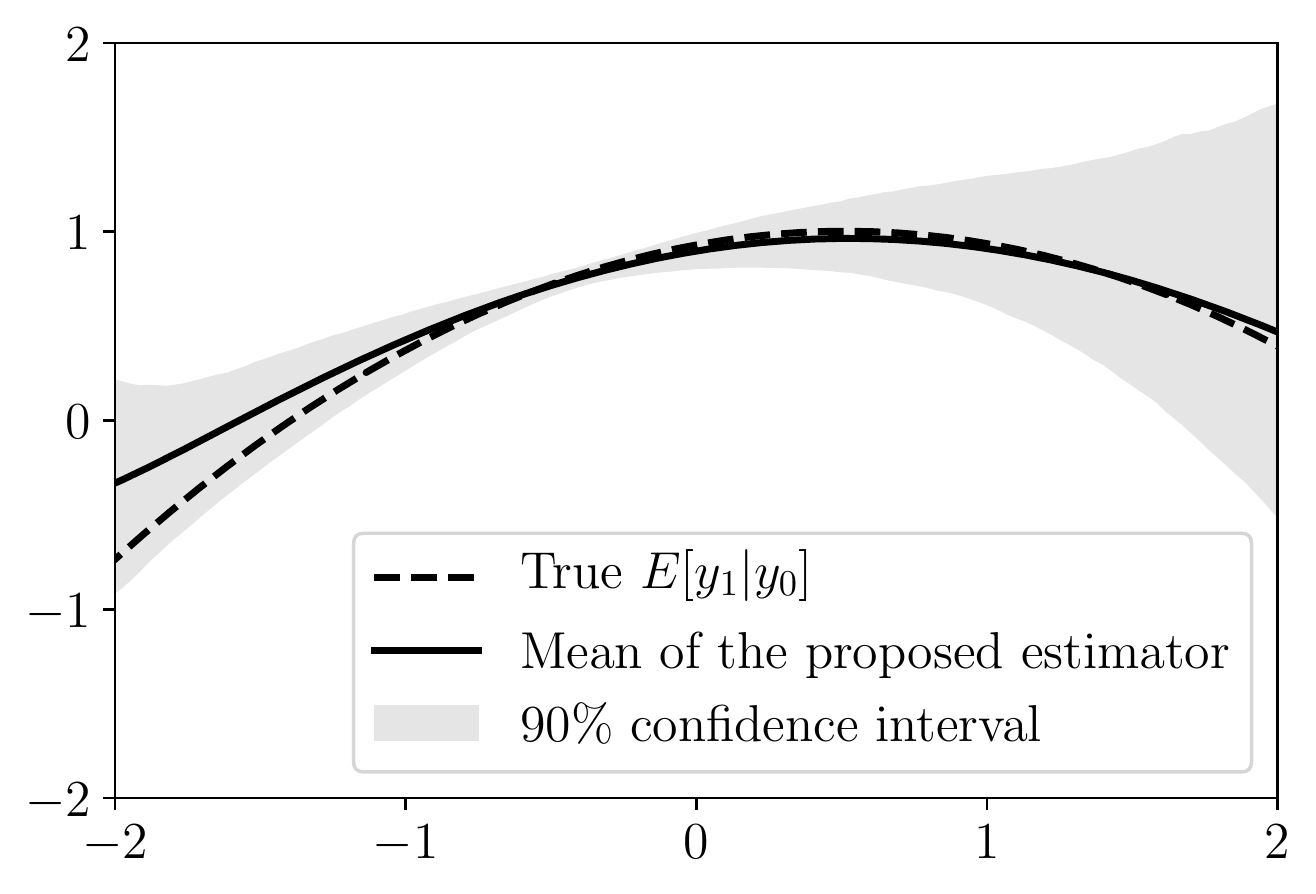}}
	
	\caption{The result of estimation of $\Ec{y_{1}}{y_{0}}$ (horizontal axis: $y_{0}$, vertical axis: $y_{1}$)}
	\label{fig:estimation_result} 
\end{figure}

Figure \ref{fig:estimation_result} shows the result of the estimation of $\Ec{y_{1}}{y_{0}}$. The horizontal axis is $y_{0}$ and the vertical axis is $y_{1}$. The dashed line shows the theoretical value of $\Ec{y_{1}}{y_{0}}$. The solid line and the gray region show the mean of the estimator and the  90\% confidence interval from 1000 replications respectively. As the figure  shows, the performance of the estimator is substantially influenced by $B_{\gamma}$; if we set a small value as $B_{\gamma}$, the variance of the estimator also becomes small, whereas the confidence interval may not include the true curve and the expectation of the estimator may be apart from it. This problem is particularly serious on the edge, where only a small number of samples is observed. However, it is notable that we can estimate $\Ec{y_{1}}{y_{0}}$ to some extent although none of the pair $(y_{i1},y_{i0})$ is observed and there are large overlaps in the distributions (see Figure \ref{fig:plots_xy_comp}). 

Table \ref{tab:ate} shows the result of estimating ATE by integrating out $y_{0}$ in $\hat{E}[y_{1}|y_{0}]$. The theoretical value of the ATE in our setting is 0.900. As the estimation of $\Ec{y_{1}}{y_{0}}$, the variance becomes larger as $B_{\gamma}$ becomes larger. On the other hand, the mean of the estimator for the ATE gets closer to the true value when $B_{\gamma}$ is small. Although we recognize its importance, how to determine $B_{\gamma}$ is beyond the scope of this paper. 
\begin{table}[htb]
  \centering
  \caption{Estimation of ATE (true: 0.900)}
   \begin{tabular}{ccc} \hline
    $B_{\gamma}$ & mean & s.d. \\ \hline
    10 & 0.888 & 0.0243 \\
    15 & 0.884 & 0.0260 \\
    25 & 0.881 & 0.0298 \\
    50 & 0.877 & 0.0397 \\\hline
  \end{tabular}
  \label{tab:ate}
\end{table}

\section{Concluding remarks}
We proposed a semiparametric two-stage least square estimator for the HTE and examine its properties through a simple simulation study, showing the availability of estimating $\Ec{y_{1}}{y_{0}}$ even though none of the pair $(y_{1},y_{0})$ is observed. As mentioned in Section 4, the performance of the estimator shown in this paper is influenced substantially by the constraint parameter, which we have to tune. In addition, although we use Legendre polynomials up to the third-order in the simulation, the order actually needs to be determined reflecting the characteristics of the target population. Although there is literature on this issue (e.g., \cite{horowitz_adaptive_2014}), a decisive method has not been developed. More importantly, our approach would not work for a multivariate case because it uses the kernel density estimator, that is, our method suffers from the curse of dimensionality. Similarly, an approximation using the orthogonal basis functions as eq.~(\ref{eq:approx}) would be a problem because the number of parameters grows with $J^{d}$ (``approximation order" to the power ``the number of dimensions of covariates") rate. Considering this issue, a nonparametric Bayesian approach may be helpful. We are planning to address this in future work.

\bibliography{references.bib}
\bibliographystyle{cje}

\end{document}